\documentclass{svproc}
\usepackage{url}
\usepackage{graphicx}
\usepackage[colorlinks = true, citecolor = blue, urlcolor= blue]{hyperref}
\usepackage{svg}
\usepackage{cite}
\usepackage{tikz}
\usepackage{algorithm2e}
\usepackage{enumitem}
\usepackage{pifont}
\usetikzlibrary{matrix,chains,positioning,decorations.pathreplacing,arrows}
\usepackage{subfigure}
\usepackage{amsmath}
\usepackage{multirow}
\usepackage{xcolor}
\usepackage{xspace}
\SetCommentSty{mycommfont}

\newcommand{\orcid}[1]{\href{https://orcid.org/#1}{\includesvg[width=10pt]{orcid}}}

\begin{document}
\mainmatter              % start of a contribution
\title{A Dynamic Weighted Federated Learning for Android Malware Classification}

\author{Ayushi Chaudhuri$^{1}$\orcid{0000-0002-0001-3202}, Arijit Nandi$^{2,3}$\orcid{0000-0003-4238-5183} \and Buddhadeb Pradhan$^{4}$\orcid{0000-0003-3496-3625} }
\authorrunning{Chaudhuri et al.} % abbreviated author list (for running head)
% \tocauthor{Ivar Ekeland, Roger Temam, Jeffrey Dean, David Grove,
% Craig Chambers, Kim B. Bruce, and Elisa Bertino}
%
\institute{$^{1}$Dept. of Computer Science and Engineering, Vellore Institute of Technolog(VIT) Bhopal, India,\\
$^{2}$Department of Computer Science, Universitat Politècnica de Catalunya (BarcelonaTech), Barcelona, Spain\\
$^{3}$Eurecat, Centre Tecnològic de Catalunya, Barcelona, Spain\\
$^{4}$Department of Computer Science and Engineering, University of Engineering and Management, Kolkata\\
\email{ayushichaudhuri10@gmail.com (A.C), jit.ari172@gmail.com (A.N), buddhadebpradhan@gmail.com (B.P)}}

\maketitle              % typeset the title of the contribution

\begin{abstract}
Android malware attacks are increasing daily at a tremendous volume, making Android users more vulnerable to cyber-attacks. Researchers have developed many machine learning (ML)/ deep learning (DL) techniques to detect and mitigate android malware attacks. However, due to technological advancement, there is a rise in android mobile devices. Furthermore, the devices are geographically dispersed, resulting in distributed data. In such scenario, traditional ML/DL techniques are infeasible since all of these approaches require the data to be kept in a central system; this may provide a problem for user privacy because of the massive proliferation of Android mobile devices; putting the data in a central system creates an overhead. Also, the traditional ML/DL-based android malware classification techniques are not scalable. Researchers have proposed federated learning (FL) based android malware classification system to solve the privacy preservation and scalability with high classification performance. In traditional FL, Federated Averaging (FedAvg) is utilized to construct the global model at each round by merging all of the local models obtained from all of the customers that participated in the FL. However, the conventional FedAvg has a disadvantage: if one poor-performing local model is included in global model development for each round, it may result in an under-performing global model. Because FedAvg favors all local models equally when averaging. To address this issue, our main objective in this work is to design a dynamic weighted federated averaging (DW-FedAvg) strategy in which the weights for each local model are automatically updated based on their performance at the client. The DW-FedAvg is evaluated using four popular benchmark datasets, Melgenome, Drebin, Kronodroid and Tuandromd used in android malware classification research. The results show that our proposed approach is scalable, privacy preserved, and capable of outperforming the traditional FedAvg for android malware classification in terms of accuracy, F1 score, AUC score and FPR score. 
\keywords{Android Malware Classification, Federated Learning, Android Security, Distributed Machine Learning, Artificial Neural Network}
\end{abstract}
\section{Introduction}
Now a days, android has become one of the most widely and popularly used operating system~\cite{kevin2014}. Also, the threats (android malware) in the android operating system have increased at a rapid rate. It has been found that the share of android malware is higher than 46\% among all types of mobile malwares and 400\% increase in android-based malware since 2010~\cite{zohu2012}. Malware is a type of malicious software that targets the mobile devices running on android operating system. Android devices now have new features for installing and using apps compared to traditional computers. Because of this new feature there is a chance that the android device become vulnerable to android malwares~\cite{Android2016} and it could be even difficult to safe guard the devices properly from the malwares. Malware infected devices posses serious threats to not only the users but also an organization.

With the advancement of machine learning (ML) or deep learning (DL) algorithms their application in android malware classification is increasing day by day due to its effectiveness~\cite{Droidfusion}. Furthermore, traditional ML/DL based malware classification techniques are not suitable and scalable in the current scenario (rapid growth of mobile devices) because of the following reasons: 
\begin{enumerate}
    \item {Decentralised data: users are geographically located so generated data is distributed.}
    \item {Data have sensitive information  such as location data, online identifiers (IP address) etc.}
\end{enumerate}

Federated cybersecurity is one of the newest and emerging approach in malware detection and classification~\cite{Bimal2022}. It makes the detection of cyber threats more secure and also use the IoT network system efficiently. This paper focuses on a detailed study of federated models for cybersecurity and machine learning by dividing them into two parts. The first one describes the FL and how it can be applied in cybersecurity for IoT and the second part addresses cybersecurity for FL. This survey mainly focuses on security approaches and also give importance to performance issues related to FL. Security attacks and preventive measures are summarized and also performance issues in FL for IoT networks are also described proficiently. In~\cite{Fed-IIoT2020}, authors have proposed a federated-learning (FL) architecture along with android malware detection algorithm, known as Fed-IIoT. This architecture consists of two parts, first one is participant side, where the data has been triggered by two dynamic poisoning attacks based on generative adversarial network (GAN) and federated GAN; and the second one is server side, which monitors the global model and gives a shape to the robust collaboration training model. This model proposes to avoid anomaly in aggregation by a GAN network defense algorithm to detect the vulnerabilities in the server side and also adjusts and adapts Byzantine defense algorithm on Krum and Medium to increase its effectiveness. By using Fed-IIoT, devices can safely communicate with each other with no privacy issues. These features are employed to classify different malwares by using convolutional neural networks (CNN).  Similar kind of research found in~\cite{Perm2022}, where a new technique called permission maps is developed, which provides combined information of android permissions and their severity levels. The training phase of the Perm-Maps is supported by a federated architecture. A CNN model then is employed to classify different malware families. At last, a feature selection approach is applied to reduce the computational effort and CNN training processes. Authors in~\cite{Lim2021} has developed LiM ('Less is More'), which is a malware classification framework that uses federated learning to detect and categorise dangerous programmes by protecting the privacy of others. LiM employs a safe semi-supervised ensemble learning and FL that maximises malware classification accuracy. Researchers have proposed federated learning based approaches for android malware classification~\cite{Fed-IIoT2020}. In traditional FL, Federated Averaging (FedAvg) is utilized to construct the global model at each round by merging all of the local models obtained from all of the customers that participated in the FL. However, the conventional FedAvg has a disadvantage: if one poor-performing local model is included in global model development for each round, it may result in an under-performing global model. Because FedAvg favors all local models equally when averaging.

To solve this above mentioned issues, we propose a dynamic weighted federated averaging (DW-FedAvg) strategy in which the weights for each local model are automatically updated based on their performance at the client to classify android malware classification. The main contributions of our paper are as follows: 

\begin{itemize}
    \item{We a present a dynamic weighted federated averaging (DW-FedAvg) strategy in federated learning framework for android malware classification.}
    \item {The proposed DW-FedAvg is capable of delivering a high accuracy global classifier without accessing the distributed data while classifying android malwares.}
    \item {The experimental results based on the popular Drebin, Malgenome, Kronodroid and Tuandromd data sets showed that the DW-FedAvg achieves high accuracy, it is scalable.}
\end{itemize}

The rest of the paper is structured as follows: brief introduction to the ideas of android malware and federate learning are presented in Section~\ref{preli}. In Section~\ref{materials-methods}, the material and methods for DW-FedAvg is presented and in Section~\ref{results-analysis}, the experiment results and discussion is provided. Finally the paper ends with the conclusion in Section~\ref{conclusion}.

\section{Preliminaries}\label{preli}
Here in this section we have provided the brief introduction about android malware and federated learning. 
\subsection{Android Malware}
Android malware is a type of malicious software that targets the mobile devices running on android operating system. It has been growing at a significant rate. It has been found that the share of android malware is higher than 46\% than that among all types of mobile malwares~\cite{zohu2012}. There is also 400\% increase in android-based malware since 2010~\cite{zohu2012}. Android devices now have new features for installing and using apps compared to traditional computers. This new feature makes it even difficult for the android operating system to defend against android malware~\cite{Android2016}. 
\subsection{Federated Learning}
Federated Learning (FL~\cite{FedL,ZHANG2022171}) is a promising distributed machine learning approach that enables mobile local clients to build a powerful global model by collaborating with a global server. The mobile devices share a local model developed from the sensitive data accessible to those devices without sharing the sensitive data with other parties. Clearly, in FL there are two parties involved in the whole approach, global server and local clients (end users or edge user devices)~\cite{FedL}. 

The main functions of local clients are: (1) access to the local data, develop the model and perform the corresponding learning task (2) sharing the developed local model to the global server (3) receive the global model and replace the local model with the global model.  

The main functions of the global server are: (1) collect the shared local models from local clients; (2) create the global model by model averaging (called Federated Averaging) of all the collected local models; (3) broadcast the global model to all the local models participated in the Federated Averaging (FedAvg). 

\begin{figure}[htbp]
    \centering
    \includegraphics[width=0.6\textwidth]{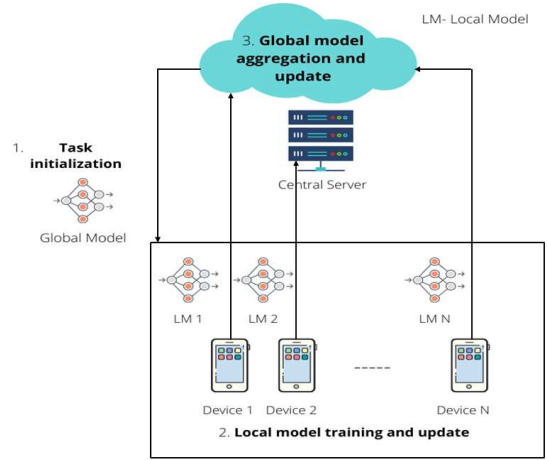}
    \caption{The general framework of federated learning}
    \label{fig:FL}
\end{figure}

In federated learning the global server and local clients continues performing the above mentioned functions until a desirable model accuracy is achieved~\cite{Pan2021}. In Fig.~\ref{fig:FL} the general framework of FL is presented.

\section{Materials and Methods}\label{materials-methods}
In this section, the proposed approach details, the benchmark dataset description, experimental setup, and finally the performance metric are provided. 
\subsection{Proposed Approach}
At each round of FL, a global model is created by averaging (Federated Averaging (FedAvg)) all of the local models received from all of the clients that participated in the FL. However, the traditional FedAvg has a drawback: such as if one poor-performing local model is included in global model construction for each round, it might result in an under-performing global model. Because FedAvg prioritizes all local models equally while averaging. In this scenario, we can use the weighted averaging approach, in which a weight is assigned to each of the local models. The weight assigned to each local model is very challenging because it's done by trial and error and the overall performance depends on the weight assignment. So, to solve this problem, in this paper, our main motivation is to develop a dynamic weighted federated averaging (DW-FedAvg) approach where the weights for each local model are adjusted automatically based on their performance at the client.
The FedAvg is performed at global server based on the following Eq:~\ref{eq:fedavg} as follows: 
\begin{equation}
    w^{g}_{t} = \frac{1}{N_{c}}\sum_{1}^{N_{c}}w^{l}_{t-1,i} 
    \label{eq:fedavg}
\end{equation}

Now, the DW-FedAvg is performed based on the Eq.~\ref{eq:dw-fedavg} at global server: 
\begin{equation}
    w^{g}_{t} = \frac{1}{N_{c}}\sum_{1}^{N_{c}}\beta_i \cdot w^{l}_{t-1,i} 
    \label{eq:dw-fedavg}
\end{equation}

\noindent Where the $w^{g}_t$ is the global model created at time $t$, $N_c$ is the total number of local model received at global server, it also indicates total number of clients participated in the FL, $w^{l}_{t-1,i}$ is the local model received from all the clients at time $t-1$ and the $\beta$ is the dynamic weights associated with each local models rceieved. In this paper, we have considered full participation from all the clients in the FedAvg and DW-FedAvg. 

The dynamic weight $\beta$ is adjusted automatically based on each local model's performance at client's end. To do so, the global server first consider each local model has equal powerful, hence equal priority. The global server creates the a global priority index matrix which contains the weights corresponding to each local models. Then the weights are adjusted automatically based on the local models performance. The dynamic weight changes are based on the following condition:

\begin{itemize}
    \item {\textbf{Initialize:} $\beta_{i,t} = \frac{1}{N_c}$, $\alpha (the\; reward/penalty\; factor) = 0.2$.} 
    \item {$Acc_p \gets 0$ and $Acc_c \gets$ all local model's test accuracy.}
    \item{If \; round == 1 then $Acc_p = Acc_c$.}
    \item {Else:
        \begin{itemize}
            \item {If ${Acc_c}_{i} > {Acc_p}_{i}$ then $\beta_{t,i} = \beta_{t,i}+\beta_{t,i}*\alpha$.}
            \item{ElseIf ${Acc_c}_{i} < {Acc_p}_{i}$ then $\beta_{t,i} = \beta_{t,i}-\beta_{t,i}*\alpha$.}
            \item{Else do nothing.}
        \end{itemize}
    }
    \item{Weight re-scaling: $\beta_{t,i} = \frac{\beta_{t,i}}{sum(\beta_{t,i})}$ }
    \item {repeat everything till all rounds end.}
\end{itemize}

\subsection{Dataset Description}
We have used four datasets for our approach which are publicly available. The brief description of those considered datasets are as follows: 
\begin{enumerate}
    \item {\textbf{Malgenome:} This dataset contains features from 3799 app samples where 2539 are benign and 1260 are android malwares from Android malware genome project~\cite{Droidfusion}. It contains total features of 215. }
    \item {\textbf{Drebin:}This dataset contains features from 15036 app samples where 9476 are benign and 5560 are android malwares from Drebin project~\cite{ZHANG2022171}. It also contains 215 features.}
    \item {\textbf{Tunadromd:} This dataset~\cite{Tunadromd} contains features from 4465 app samples where 903 are benign and 3565 are android malwares. It contains total features of 241.}
    
    \item{\textbf{Kronodrid:} This dataset contains features from 78137 app samples where 36935 are benign and 41382 are android malwares~\cite{Kronodroid}. It contains total features of 463.}
\end{enumerate}

The brief description of those considered benchmark datasets are presented in TABLE~\ref{tab:dataset-descrption}.

\begin{table}[htbp]
\centering
\caption{Dataset description}\label{tab:dataset-descrption}
\resizebox{0.5\linewidth}{!}{\begin{tabular}{|c|c|c|c|}
\hline
\textbf{\begin{tabular}[c]{@{}c@{}}Dataset \\ Name\end{tabular}} & \textbf{\begin{tabular}[c]{@{}c@{}}No. of \\ Samples\end{tabular}} & \textbf{\begin{tabular}[c]{@{}c@{}}No. of\\ attributes/features\end{tabular}} & \textbf{\begin{tabular}[c]{@{}c@{}}Class\\ labels\end{tabular}} \\ \hline
Malgenome & 3799 & 215 & \begin{tabular}[c]{@{}c@{}}Benign (2539)\\ Malware (1260)\end{tabular} \\ \hline
Drebin & 15036 & 215 & \begin{tabular}[c]{@{}c@{}}Benign (9476)\\ Malware (5560)\end{tabular} \\ \hline
Tunadromd & 4465 & 241 & \begin{tabular}[c]{@{}c@{}}Benign (903)\\ Malware (3565)\end{tabular} \\ \hline
Kronodrid & 78137 & 463 & \begin{tabular}[c]{@{}c@{}}Benign (36935)\\ Malware (41382)\end{tabular} \\ \hline
\end{tabular}}
\end{table}

\subsection{Experimental setup}
Herewith the machine setup, software development, experimental environment and parameter setup are as follows:

\begin{itemize}
    \item  \textit{Machine configuration:} Ubuntu 20.04 64 bit OS, processor core-i7-7700HQ with RAM 24 Gb--2400MHz and 4Gb-Nvidia GTX-1050 graphics.

    \item  \textit{Software development:} The DW-FedAvg is implemented in Python~3.7 with the help of TensorFlow 2.0 and Keras in the backend.
    
    \item  \textit{Base classifier and optimizer:} For the base classifier we have used 4-layer feed-forward network with 1st hidden layer contains 200 neurons 2nd hidden layer contains 100 and the 3rd hidden layer contains 50 neurons. The selection of layers is done by trail-error approach because there is approach to set it automatically. In hidden layers ReLU is the activation function and in the output layer sigmoid is the activation function. In the output layer sigmoid being the activation function is because we have considered binary classification (Benign or Malware). For training the neural network we have used Stochastic Gradient Descent (SGD) optimizer because its popularly used in federated approach. 
    \item  \textit{Parameter setup:} The batch size of the DW-FedAvd is 32, the epoche is 32. We have used 80-20 Hold-Out cross-validation technique to train and test the model and the overall performance comparison. The learning rate of SGD is 0.01.

    \item \textit{Source-code:}  The source code and implementation details of our proposed approach can be found in \textit{Github}\footnote{The source code can be found in GitHub at: \url{https://github.com/officialarijit/DW-FedAvg}}.

\end{itemize}

\subsection{Performance Metric}

The accuracy, F1-score, area under the ROC curve (AUC) and False Positive Rate (FLR) are metrices for evaluating classification model. For binary classification, the mathematical formulae are as calculated in terms of positives and negatives: $Accuracy=\frac{TP+TN}{TP+TN+FP+FN}$, $F1-score = \frac{2 * Precision * Recall}{Precision + Recall} = \frac{TP}{TP+\frac{1}{2}*(FP+FN)}$, $AUC = \frac{\Sigma Rank(+)-|+|*\frac{|+|+1}{2}}{|+|+|-|}$ and $  FPR = \frac{FP}{FP+TN}$. Where TP = True Positives, TN = True Negatives, FP= False Positives, and FN = False Negatives. $\Sigma Rank(+)$ is the sum the ranks of all the positively classifier samples, $|+|$ is total number of positive samples and $|-|$ is the total negative samples present in the dataset.

\section{Results, Analysis and Discussion}\label{results-analysis}
In this section, we summarize the experimental results of our proposed DW-FedAvg and make a comparison with traditional FedAvg~\cite{FedL} under different number of clients and different number of rounds. For our experiment we have considered a total of 3 different client scenario (5, 10,  and 15). Also, two different rounds are considered such as 10 and 20. The average test average test accuracy, F1-score, AUC and FPR of the global model are presented in TABLE ~\ref{tab:round-10-comp} and ~\ref{tab:round-20-comp}. The better results are in bold in the table.

\begin{table}[htbp]
\centering
\caption{Global model's average test accuracy, F1-score, AUC and FPR comparison between FedAvg and DW-FedAvg for 10 rounds}
\label{tab:round-10-comp}
\resizebox{\linewidth}{!}{\begin{tabular}{cccccccccc}
\hline
 &  & \multicolumn{4}{c}{\textbf{FedAvg}} & \multicolumn{4}{c}{\textbf{Dw FedAvg}} \\
\textbf{Dataset} & \textbf{\begin{tabular}[c]{@{}c@{}}Number of\\ Clients\end{tabular}} & \textbf{Accuracy} & \textbf{F1-score} & \textbf{AUC} & \textbf{FPR} & \textbf{Accuracy} & \textbf{F1-score} & \textbf{AUC} & \textbf{FPR} \\ \hline
\multirow{3}{*}{\textbf{Drebin}} & 5 & 0.9826$\pm$0.001 & 0.9764$\pm$0.001 & 0.9955$\pm$0.0001 & 0.0277$\pm$0.002 & \textbf{0.9828$\pm$0.001} & \textbf{0.9766$\pm$0.001} & \textbf{0.996$\pm$0.0002} & \textbf{0.0256$\pm$0.003} \\
 & 10 & 0.9812$\pm$0.002 & 0.9744$\pm$0.003 & 0.9962$\pm$0.0003 & 0.0269$\pm$0.004 & 0.9784$\pm$0.003 & 0.9705$\pm$0.005 & 0.9956$\pm$0.0004 & 0.0362$\pm$0.006 \\
 & 15 & 0.9742$\pm$0.004 & 0.9648$\pm$0.005 & 0.9947$\pm$0.001 & 0.0392$\pm$0.005 & 0.9734$\pm$0.003 & 0.9636$\pm$0.005 & 0.9944$\pm$0.001 & 0.0440$\pm$0.006 \\ \hline
\multirow{3}{*}{\textbf{Malgenome}} & 5 & 0.9911$\pm$0.0006 & 0.9871$\pm$0.0008 & 0.9998$\pm$0.0004 & 0.0050$\pm$0.003 & \textbf{0.9943$\pm$0.001} & \textbf{0.9917$\pm$0.001} & \textbf{0.9999$\pm$0.0003} & \textbf{0.0050$\pm$0.003} \\
 & 10 & 0.9901$\pm$0.003 & 0.9855$\pm$0.005 & 0.9994$\pm$0.0002 & 0.0135$\pm$0.011 & 0.9896$\pm$0.003 & 0.9847$\pm$0.004 & 0.9990$\pm$0.0006 & 0.0154$\pm$0.010 \\
 & 15 & 0.9838$\pm$0.001 & 0.9763$\pm$0.001 & 0.9988$\pm$0.0006 & 0.0189$\pm$0.005 & \textbf{0.9875$\pm$0.005} & \textbf{0.9814$\pm$0.008} & \textbf{0.9993$\pm$0.0003} & \textbf{0.0231$\pm$0.018} \\ \hline
\multirow{3}{*}{\textbf{Kronodroid}} & 5 & 0.9632$\pm$0.003 & 0.9651$\pm$0.003 & 0.9898$\pm$0.001 & 0.0391$\pm$0.005 & 0.9596$\pm$0.004 & 0.9618$\pm$0.004 & 0.9888$\pm$0.001 & 0.0420$\pm$0.005 \\
 & 10 & 0.9548$\pm$0.003 & 0.9569$\pm$0.002 & 0.9870$\pm$0.001 & 0.0514$\pm$0.002 & 0.9516$\pm$0.003 & 0.9538$\pm$0.002 & 0.9855$\pm$0.001 & 0.0567$\pm$0.002 \\
 & 15 & 0.9498$\pm$0.004 & 0.9521$\pm$0.003 & 0.9856$\pm$0.001 & 0.0566$\pm$0.003 & 0.9478$\pm$0.004 & 0.9502$\pm$0.003 & 0.9844$\pm$0.002 & 0.0597$\pm$0.003 \\ \hline
\multirow{3}{*}{\textbf{Tuandromd}} & 5 & 0.9880$\pm$0.002 & 0.9926$\pm$0.001 & 0.9988$\pm$0.0004 & 0.0064$\pm$0.002 & 0.9861$\pm$0.002 & 0.9914$\pm$0.001 & 0.9971$\pm$0.0003 & 0.0093$\pm$0.001 \\
 & 10 & 0.9840$\pm$0.004 & 0.9902$\pm$0.002 & 0.9983$\pm$0.0004 & 0.0133$\pm$0.004 & \textbf{0.9857$\pm$0.005} & \textbf{0.9912$\pm$0.003} & \textbf{0.9985$\pm$0.0009} & \textbf{0.0126$\pm$0.005} \\
 & 15 & 0.9799$\pm$0.008 & 0.9876$\pm$0.004 & 0.9977$\pm$0.001 & 0.0149$\pm$0.006 & 0.9780$\pm$0.003 & 0.9864$\pm$0.002 & 0.9967$\pm$0.001 & 0.0215$\pm$0.001 \\ \hline
\end{tabular}}
\end{table}

\begin{table}[htbp]
\centering
\caption{Global model's average accuracy, F1-score, AUC and FPR comparison between FedAvg and DW-FedAvg for 20 rounds}
\label{tab:round-20-comp}
\resizebox{\linewidth}{!}{\begin{tabular}{cccccccccc}
\hline
\textbf{} & \textbf{} & \multicolumn{4}{c}{\textbf{FedAvg}} & \multicolumn{4}{c}{\textbf{DW FedAvg}} \\
\textbf{Dataset} & \textbf{Number of clients} & \textbf{Accuracy} & \textbf{F1-score} & \textbf{AUC} & \textbf{FPR} & \textbf{Accuracy} & \textbf{F1-score} & \textbf{AUC} & \textbf{FPR} \\ \hline
\multirow{3}{*}{\textbf{Drebin}} & \textbf{5} & 0.9848$\pm$0.001 & 0.9793$\pm$0.001 & 0.9962$\pm$0.0009 & 0.0232$\pm$0.001 & 0.9841$\pm$0.002 & 0.9783$\pm$0.003 & 0.9961$\pm$0.0002 & 0.0268$\pm$0.003 \\
 & \textbf{10} & 0.9808$\pm$0.001 & 0.9740$\pm$0.002 & 0.9957$\pm$0.003 & 0.0258$\pm$0.004 & \textbf{0.9825$\pm$0.003} & \textbf{0.9763$\pm$0.004} & 0.9957$\pm$0.0003 & 0.0268$\pm$0.005 \\
 & \textbf{15} & 0.9794$\pm$0.002 & 0.9720$\pm$0.003 & 0.9958$\pm$0.0003 & 0.0314$\pm$0.002 & 0.9780$\pm$0.005 & 0.9701$\pm$0.006 & 0.9946$\pm$0.0005 & 0.0337$\pm$0.008 \\ \hline
\multirow{3}{*}{\textbf{Malgenome}} & \textbf{5} & 0.9962$\pm$0.001 & 0.9944$\pm$0.002 & 0.9997$\pm$0.0002 & 0.0054$\pm$0.003 & 0.9923$\pm$0.001 & 0.9887$\pm$0.002 & 0.9998$\pm$0.0001 & 0.0050$\pm$0.004 \\
 & \textbf{10} & 0.9892$\pm$0.001 & 0.9842$\pm$0.002 & 0.9994$\pm$0.0002 & 0.0090$\pm$0.006 & \textbf{0.9940$\pm$0.003} & \textbf{0.9912$\pm$0.005} & 0.9998$\pm$0.0001 & 0.0079$\pm$0.009 \\
 & \textbf{15} & 0.9921$\pm$0.002 & 0.9884$\pm$0.003 & 0.9997$\pm$0.001 & 0.0088$\pm$0.008 & 0.9897$\pm$0.004 & 0.9849$\pm$0.006 & 0.9992$\pm$0.0007 & 0.0137$\pm$0.013 \\ \hline
\multirow{3}{*}{\textbf{Kronodroid}} & \textbf{5} & 0.9683$\pm$0.002 & 0.9700$\pm$0.002 & 0.9909$\pm$0.0007 & 0.0321$\pm$0.004 & 0.9661$\pm$0.003 & 0.9680$\pm$0.003 & 0.9904$\pm$0.001 & 0.0331$\pm$0.003 \\
 & \textbf{10} & 0.9623$\pm$0.002 & 0.9644$\pm$0.002 & 0.9902$\pm$0.001 & 0.0372$\pm$0.003 & 0.9622$\pm$0.003 & 0.9643$\pm$0.003 & 0.9896$\pm$0.001 & 0.0385$\pm$0.005 \\
 & \textbf{15} & 0.9607$\pm$0.004 & 0.9627$\pm$0.004 & 0.9884$\pm$0.001 & 0.0405$\pm$0.007 & 0.9598$\pm$0.005 & 0.9619$\pm$0.005 & 0.9881$\pm$0.001 & 0.0432$\pm$0.007 \\ \hline
\multirow{3}{*}{\textbf{Tuandromd}} & \textbf{5} & 0.9893$\pm$0.001 & 0.9934$\pm$0.001 & 0.9992$\pm$0.0002 & 0.0050$\pm$0.001 & 0.9870$\pm$0.003 & 0.9920$\pm$0.002 & 0.9975$\pm$0.0005 & 0.00961$\pm$0.002 \\
 & \textbf{10} & 0.9843$\pm$0.002 & 0.9904$\pm$0.001 & 0.9987$\pm$0.0003 & 0.0100$\pm$0.002 & \textbf{0.9867$\pm$0.005} & \textbf{0.9918$\pm$0.003} & 0.9989$\pm$0.0006 & 0.0093$\pm$0.004 \\
 & \textbf{15} & 0.9839$\pm$0.006 & 0.9901$\pm$0.003 & 0.9960$\pm$0.0008 & 0.0134$\pm$0.005 & 0.9832$\pm$0.005 & 0.9896$\pm$0.003 & 0.9981$\pm$0.0008 & 0.0130$\pm$0.005 \\ \hline
\end{tabular}}
\end{table}

For the Malgenome dataset our proposed DW-FedAvg approach has shown almost similar score or greater score in all aspect (accuracy, F1 score, auc score and FPR score) compare to the traditional FedAvg approach. In the first comparison~\ref{tab:round-10-comp}, (where no. of rounds is 10) there are 5 clients. Our proposed DW-FedAvg approach provides an increased accuracy and F1 score of 0.02\% and an increased auc score of 0.05\% and a decreased FPR score of 0.21\% than the FedAvg approach for those first 5 clients. For 10 clients, it gives almost similar accuracy, F1 score and auc score as that of FedAvg approach, whereas the FPR score has been increased to 0.93\%. Likewise for 15 clients, it gives a slight decrease or almost similar accuuracy, F1 score and auc score as that of FedAvg approach, whereas the FPR score has been increased to 0.48\%.  Similarly for round 20~\ref{tab:round-20-comp}, it is also tested among 5, 10 and 20 clients. For 5 clients, it gives almost similar accuracy, F1 score and auc score as that of FedAvg approach, whereas the FPR score has been increased to 0.36\%. For 10 clients, our proposed approach provides an increased accuracy of 0.17\% and F1 score of 0.23\% and an increased FPR score of 0.1\% than the FedAvg approach. For 15 clients, it gives a slight decrease or almost similar accuracy, F1 score and auc score as that of FedAvg approach, while the FPR score has increased to 0.23\%.

Similarly for the Drebin dataset, our DW-FedAvg approach also provides almost similar score or greater score in all aspect compare to the traditional FedAvg approach. This dataset has also been tested through two rounds-10 and 20. In round 10 ~\ref{tab:round-10-comp}, it is tested among 5, 10 and 20 clients. For 5 clients, the model provides a significant increase of accuracy of 0.32\%, F1 score of 0.46\% and auc score of 0.01\% as that of FedAvg approach. For 10 clients, our proposed approach gives almost similar accuracy, F1 score and auc score as that of FedAvg approach, whereas the FPR score has increased to 0.19\%. For 15 clients, our DW-FedAvg approach provides an increased accuracy of 0.37\%, F1 score of 0.51\%, auc score of 0.05\% and an FPR score of 0.42\% than the FedAvg approach. Similarly for round 20 ~\ref{tab:round-20-comp}, it is tested among 5, 10 and 20 clients. For 10 clients, a significant increase of accuracy score of 0.48\%, F1 score of 0.7\% and auc score of 0.04\% is observed in our approach than the FedAvg approach. For 5 and 15 clients, it gives a slight decrease or almost same accuracy, F1 score, auc score and FPR score as that of FedAvg approach.

For the Kronodroid dataset, our proposed DW-FedAvg approach provides almost similar score or greater score as well in all aspect i.e., accuracy, F1 score, auc score and FPR score, in compare to the traditional FedAvg approach. Likewise, the other datasets, this dataset has also been tested through two rounds-10 and 20. In both the rounds, it is tested among 5, 10 and 20 clients. In round 10 ~\ref{tab:round-10-comp}, for 5 and 10 clients, our proposed approach gives a slight decrease or almost similar accuracy, F1 score, auc score and FPR score as that of FedAvg approach. For 15 clients, our proposed approach gives almost similar accuracy, F1 score and the auc score, whereas there is a significant increase of the FPR score of 0.31\% than the FedAvg approach. In round 20 ~\ref{tab:round-20-comp}, for 20 clients, our proposed approach gives almost similar accuracy, F1 score, auc score and FPR score as that of FedAvg approach. For 5 and 10 clients, the accuracy, F1 score and the auc score of our approach almost remains same as that of FedAvg approach, while the FPR score has been increased to 0.1\% and 0.13\% respectively than the FedAvg approach.
Likewise, for Tuandromd dataset, our DW-FedAvg approach also provides almost similar score or greater score in all aspect in compare to the traditional FedAvg approach. This dataset has also been tested through two rounds-10 and 20. In round 10 ~\ref{tab:round-10-comp}, it is tested among 5, 10 and 20 clients. For 5 and 15 clients, our approach provides almost similar accuracy, F1 score, auc score and FPR score to that of FedAvg approach. For 10 clients, there is a significant increase of accuracy by 0.17\%, F1 score by 0.1\% and auc score by 0.02\% and a slight decrease of FPR score by 0.07\%. In round 20 ~\ref{tab:round-20-comp}, for 5 clients, our approach provides almost similar accuracy, F1 score, auc score and FPR score to that of FedAvg approach. For 10 clients, our DW-FedAvg approach provides a significant increase of the accuracy by 0.24\%, F1 score by 0.14\% and auc score by 0.02\% and a slight decrease of FPR score by 0.07\% than the FedAvg approach. For 20 clients, our approach provides an increase in auc score by 0.21\%, whereas the accuracy, F1 score and the FPR score remains almost same as that of FedAvg approach.

% The reason why the FedAvg is better than our proposed approach in some cases?
The FedAvg approach takes the average of the local models. Our DW-FedAvg approach adjusts the weight based on the model performance. Our approach rewards best performing models whereas penalizes the poor performing models. In case of too many local models with poor classifiers, our DW-FedAvg approach dynamically penalizes weights based on their performance, and thus minimum number of models are getting rewarded. Therefore, the dynamic average overall decreases the accuracy of the model. On the other hand, traditional FedAvg performs average on the local models and does not dynamically adjust the weights of the local models, thus it does not have any effect on accuracy degradation.   

% The reason why DW-FedAvg is better than traditional FedAvg?
In traditional FedAvg approach, simple averaging of the local models is done, and it results into a global model. Thus, equal priority is given to all the models. On the other hand, our DW-FedAvg approach dynamically adjusts the weight of the local models based on their performance. Our approach penalizes the poor performing classifiers, whereas rewards the best performing classifiers. By doing the simple average can create a bad global model for the fact that if a global model performs poorly. Our approach prioritizes the best performing models. As more priority is given to better performing models, the global model gives a better result in the accuracy of our approach.

Finally, from the detailed comparison shows that our approach has outperformed traditional FedAvg for some clients in both the rounds for both the datasets. It also shows the advantage of dynamically adjusting the weights based on the best performing and poor performing models for both the rounds.

\section{Conclusion}\label{conclusion}
In this paper, we have proposed a dynamic weighted federated averaging (DW-FedAvg) approach where the weights for each local model are adjusted automatically based on their performance at the client to create a powerful global model in federated learning based android malware classification. Our proposed DW-FedAvg gives a reward of dynamic weightage to the best performing model and subtract the dynamic weightage from the poor performing local models. The effectiveness of our prooposed approach is evaluated using four benchmark datasets Drebin, Malgenome, Kronodroid and Tuandromd. The results show that our proposed approach has outperformed the traditional FedAvg under different number of clients and different rounds while classifying android malware's.

As a future work, we have plans to work with semi-supervised approach for classifying android malware's'. We want to integrate our approach on different machine learning algorithms other than CNN such as SVM to create a new approach called Federated SVM. Also, deploying this proposed system into real scenario and examine how will it performs.   

\bibliographystyle{splncs03}
\bibliography{bibfile}

\begin{thebibliography}{10}
\providecommand{\url}[1]{\texttt{#1}}
\providecommand{\urlprefix}{URL }

\bibitem{Android2016}
Abualola, H., Alhawai, H., Kadadha, M., Otrok, H., Mourad, A.: An android-based
  trojan spyware to study the notificationlistener service vulnerability.
  Procedia Computer Science  83,  465--471 (2016), the 7th International
  Conference on Ambient Systems, Networks and Technologies (ANT 2016) / The 6th
  International Conference on Sustainable Energy Information Technology
  (SEIT-2016) / Affiliated Workshops

\bibitem{kevin2014}
Allix, K., Jerome, Q., Bissyandé, T.F., Klein, J., State, R., Traon, Y.L.: A
  forensic analysis of android malware -- how is malware written and how it
  could be detected? In: 2014 IEEE 38th Annual Computer Software and
  Applications Conference. pp. 384--393 (2014)

\bibitem{Tunadromd}
Borah, P., Bhattacharyya, D., Kalita, J.: Malware dataset generation and
  evaluation. In: 2020 IEEE 4th Conference on Information \& Communication
  Technology (CICT). pp. 1--6. IEEE (2020)

\bibitem{Perm2022}
D’Angelo, G., Palmieri, F., Robustelli, A.: A federated approach to android
  malware classification through perm-maps. Cluster Computing  (2022)

\bibitem{Bimal2022}
Ghimire, B., Rawat, D.B.: Recent advances on federated learning for
  cybersecurity and cybersecurity for federated learning for internet of
  things. IEEE Internet of Things Journal  9(11),  8229--8249 (2022)

\bibitem{Kronodroid}
Guerra-Manzanares, A., Bahsi, H., Nõmm, S.: Kronodroid: Time-based
  hybrid-featured dataset for effective android malware detection and
  characterization. Computers \& Security  110,  102399 (2021)

\bibitem{Lim2021}
Gálvez, R., Moonsamy, V., Diaz, C.: Less is more: A privacy-respecting android
  malware classifier using federated learning. Proceedings on Privacy Enhancing
  Technologies  2021(4),  96--116 (2021),
  \url{https://doi.org/10.2478/popets-2021-0062}

\bibitem{FedL}
Konečný, J., McMahan, H.B., Yu, F.X., Richtarik, P., Suresh, A.T., Bacon, D.:
  Federated learning: Strategies for improving communication efficiency. In:
  NIPS Workshop on Private Multi-Party Machine Learning (2016),
  \url{https://arxiv.org/abs/1610.05492}

\bibitem{Fed-IIoT2020}
Taheri, R., Shojafar, M., Alazab, M., Tafazolli, R.: Fed-iiot: A robust
  federated malware detection architecture in industrial iot. IEEE Transactions
  on Industrial Informatics  17(12),  8442--8452 (2021)

\bibitem{Droidfusion}
Yerima, S.Y., Sezer, S.: Droidfusion: A novel multilevel classifier fusion
  approach for android malware detection. IEEE Transactions on Cybernetics
  49(2),  453--466 (2019)

\bibitem{Pan2021}
Zhang, W., Wang, X., Zhou, P., Wu, W., Zhang, X.: Client selection for
  federated learning with non-iid data in mobile edge computing. IEEE Access
  9,  24462--24474 (2021)

\bibitem{ZHANG2022171}
Zhang, Y., Jiang, C., Yue, B., Wan, J., Guizani, M.: Information fusion for
  edge intelligence: A survey. Information Fusion  81,  171--186 (2022)

\bibitem{zohu2012}
Zhou, Y., Jiang, X.: Dissecting android malware: Characterization and
  evolution. In: 2012 IEEE Symposium on Security and Privacy. pp. 95--109
  (2012)

\end{thebibliography}

\end{document}